\documentclass{article}
\usepackage{booktabs}
\usepackage{graphicx} 
\usepackage{svg}
\usepackage{amsmath}
\usepackage{hyperref}
\usepackage{adjustbox}
\usepackage{authblk}

\title{Exploring Quantum-Enhanced Estimation of Financial Risk Metrics with Quantum RNG}
\author[2]{Emanuele Dri}
\author[1]{Achille Yomi}
\author[1]{Muthumanimaran Vetrivelan}
\author[1]{Ivàn Diego Exposito}
\author[1]{Cedric Kuassivi}
\affil[1]{\textbf{Scenario X}}
\affil[2]{LINKS Foundation}
\date{February 2025}

\begin{document}

\maketitle
\begin{abstract}
    In this paper, we present an approach for estimating significant financial metrics within risk management by utilizing quantum phenomena for random number generation. We explore Quantum-Enhanced Monte Carlo, a method that combines traditional and quantum techniques for enhanced precision through Quantum Random Numbers Generation (QRNG). The proposed methods can be based on the use of photonic phenomena or quantum processing units to generate random numbers. The results are promising, hinting at improved accuracy with the proposed methods and slightly lower estimates (both for VaR and CVaR estimation) using the quantum-based methodology.
\end{abstract}

\section{Introduction}
Value at Risk (VaR) and Conditional Value at Risk (CVaR, also known as Expected Shortfall, ES) are pivotal metrics within the broader context of risk management \cite{gestel2008}. 
Financial institutions employ these metrics to determine portfolio allocations that mitigate financial risk and comply with regulatory requirements and protocols.

In realistic contexts, it is not possible to analytically compute these quantities, thus the choice usually falls on Monte Carlo simulations, which allow for the estimation of VaR and CVaR \cite{hong2014}. 
As one might imagine, the precision required to determine these crucial quantities is very high. 
For this reason, it is not uncommon for financial institutions to simulate a large number of paths (on the order of $10^6-10^7$) for estimation purposes. 
However, a problem arises due to the limitations of classical computation: \textit{the inability to generate truly random numbers}. 
What occurs in the sampling process, which is core to Monte Carlo, is the use of pseudo random numbers that, although optimal for a limited number of simulations, begin to show patterns of correlation as the number of required samples increases. 
This can significantly impact the quality of the estimation process and reduce its final precision.

In this context, quantum technologies can offer a valuable advantage. 
Employing a quantum computer or even specific dedicated quantum hardware, it is possible to generate significant quantities of random numbers in a reasonable time frame \cite{Jennewein2000}. 
These numbers can then be used to obtain \textit{true random samples}, allowing the Monte Carlo process to scale with the number of paths while enhancing, rather than compromising, the accuracy of the final estimate.

In this paper, our aim is to illustrate a possible implementation of these techniques on a scale that is significant from an applicative perspective.

\section{Traditional computation of VaR and CVaR}
Value at Risk and Conditional Value at Risk are critical metrics in financial risk management. They are used to assess the risk of investment portfolios. This section explores two prominent methods for computing these metrics: Historical Simulation and Monte Carlo Simulation.

\subsection{Historical Simulation}
The Historical Simulation method estimates VaR using historical return data without requiring any assumptions about the return distribution. This approach relies directly on the empirical distribution of past returns. For a given confidence level $\alpha$, VaR is defined as the negative of the $\alpha$-percentile of historical returns:

\begin{equation}
\text{VaR}_{\alpha}(X) = - Q_{\alpha}(X),
\end{equation}
where $Q_{\alpha}(X)$ represents the $\alpha$-quantile of the return distribution.

CVaR, which measures the expected loss conditional on the loss exceeding the VaR, is computed as the average of losses greater than the VaR threshold:

\begin{equation}
\text{CVaR}_{\alpha}(X) = - \frac{1}{1-\alpha} \int_{\alpha}^{1} Q_{u}(X) \, du,
\end{equation}
where $Q_{u}(X)$ is the $u$-quantile of the return distribution.

\subsection{Monte Carlo Simulation}
The Monte Carlo Simulation method generates a large number of potential return scenarios by sampling from a predefined distribution, such as a normal or heavy-tailed distribution. This process allows for the estimation of VaR and CVaR even in cases with complex risk factors or non-linear dependencies.

Random number generation plays a crucial role in Monte Carlo simulation. To accurately model potential future scenarios, random samples are drawn from the chosen probability distribution of returns. This subsection provides a detailed description of the process:

\subsubsection{Step 1: Define the Distribution}
A suitable distribution $f_X(x)$ is chosen to represent the portfolio's returns. A common choice is represented by the Normal distribution: $$f_X(x) = \frac{1}{\sqrt{2\pi}\sigma} e^{-\frac{(x-\mu)^2}{2\sigma^2}},$$where $\mu$ is the mean and $\sigma$ is the standard deviation.

\subsubsection{Step 2: Generate Random Samples}
Using a random number generator (RNG), $N$ independent samples ${X_1, X_2, \dots, X_N}$ are drawn from $f_X(x)$. In practice, the inverse transform sampling method is commonly employed:
\begin{itemize}
\item Generate $u \sim U(0, 1)$, where $U(0, 1)$ is a uniform distribution.
\item Transform $u$ using the cumulative distribution function (CDF) $F_X^{-1}(u)$ of $f_X(x)$:
\begin{equation}
X = F_X^{-1}(u).
\end{equation}
\end{itemize}

\subsubsection{Step 3: Simulate Portfolio Returns}
For each sample $X_i$, the portfolio's return is calculated, incorporating correlations and weights:
\begin{equation}
R_i = \sum_{j=1}^M w_j X_{ij},
\end{equation}
where $w_j$ is the weight of asset $j$ in the portfolio, $X_{ij}$ is the simulated return of asset $j$ in scenario $i$, and $M$ is the number of assets.

\subsubsection{Step 4: Compute VaR}
Sort the $N$ portfolio returns in ascending order, $R_{(1)} \leq R_{(2)} \leq \dots \leq R_{(N)}$. The VaR is given by the $\alpha$-percentile:
\begin{equation}
\text{VaR}_{\alpha} \approx - R_{(\lfloor N \alpha \rfloor)},
\end{equation}
where $\lfloor \cdot \rfloor$ denotes the floor function.

\subsubsection{Step 5: Compute CVaR}
The CVaR is calculated as the average of returns exceeding the VaR threshold:
\begin{equation}
\text{CVaR}_{\alpha} \approx - \frac{1}{N_t} \sum{i \in \mathcal{T}} R_i,
\end{equation}
where $\mathcal{T}$ is the set of indices corresponding to $R_i$ values below $-\text{VaR}_{\alpha}$, and $N_t$ is the number of such instances.

\subsubsection{Advantages of Monte Carlo Simulation}
Monte Carlo simulation provides flexibility for modeling complex scenarios, incorporating non-linear risk factors, and handling non-normal return distributions. The accuracy of the results improves with the number of simulations $N$, but computational cost increases accordingly.

\section{Quantum Random Numbers generation}
Random numbers are crucial in various domains, including computational simulations, cryptography, and commercial applications like lotteries. 
The demand has surged with the emergence of quantum cryptography and information processing, leading to advances in random number generation methods and testing their randomness \cite{Jennewein2000}. 

Broadly, these methods are classified into \textbf{pseudo random generators}, which are algorithm-based and thus inherently deterministic, and \textbf{physical random generators}, which leverage the unpredictable behavior of physical phenomena. 
Pseudo random generators, despite being sophisticated in terms of period and randomness, fall short for certain applications requiring true unpredictability due to their deterministic internal states. 
Consequently, our implementation rules out these generators. 
Physical random generators often use chaotic systems perceived as random due to their complexity. 
Nonetheless, purely classical systems exhibit determinism over time, risking unseen external influences. 
Quantum mechanics offers a promising solution for truly random sources, as quantum decisions are fundamentally unpredictable \cite{Ma2016}. 

Among potential quantum processes, radioactive decay, while effective, poses precautions with high radioactivity levels needed for swift random signal generation. Alternatively, less technically demanding optical processes like photon beam splitting and single-photon polarization measurement provide feasible rapid sources of quantum random numbers.
Recently, the possibility of using Quantum Processing Units (QPUs) as quantum random number generators has become another interesting approach.

\subsection{QRNG for Monte Carlo}
As previously mentioned, to estimate the Value at Risk (VaR) of a portfolio of equities using the Monte Carlo method, one generates a large number of random scenarios for future asset prices based on probable distribution assumptions, often taking into account historical volatility and correlations between assets. 
These scenarios are typically produced by simulating the returns of the equities in the portfolio using random numbers derived from a chosen probability distribution, such as the normal distribution. 

The simulated returns are then applied to the current portfolio values to model potential future portfolio values over the VaR horizon. 
This large set of possible future outcomes allows for the construction of a distribution of portfolio values from which the VaR can be estimated as the specified percentile of the distribution, representing the maximum expected loss over the horizon at the given confidence level. 
Employing random numbers in this way helps capture the inherent uncertainties and market dynamics, providing a robust probabilistic estimate of risk.

\subsection{Quantum for randomness}
The fundamental requirement is access to quantum hardware capable of generating true random samples. These samples can then be converted from uniformly distributed samples to samples distributed according to a standard normal distribution.

It is important to note that the generation of these numbers could be performed in advance, with the results stored to be readily available for user request. This approach eliminates the need for (costly) continuous access to quantum hardware.

The first source of randomness we tested was provided by the Australian National University (ANU) \cite{anu_url}. 
Their website offers true random numbers to anyone on the internet and the methodology is described on the same website as such: 

\textit{"The random numbers are generated in real-time in our lab by measuring the quantum fluctuations of the vacuum. 
The vacuum is described very differently in the quantum physics and classical physics. 
In classical physics, a vacuum is considered as a space that is empty of matter or photons. 
Quantum physics however says that that same space resembles a sea of virtual particles appearing and disappearing all the time. This is because the vacuum still possesses a zero-point energy. 
Consequently, the electromagnetic field of the vacuum exhibits random fluctuations in phase and amplitude at all frequencies. By carefully measuring these fluctuations, we are able to generate ultra-high bandwidth random numbers."}

\begin{figure}[!ht]
    \centering
    \includegraphics[width=0.3\linewidth]{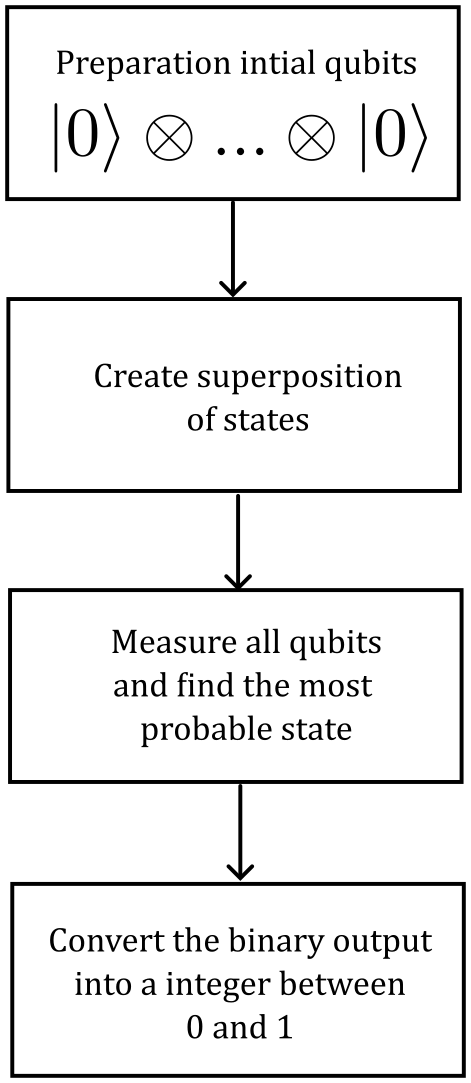}
    \caption{The flow chart of a Quantum Random Number Generator that exploits a single QPU.}
    \label{mc_4}
\end{figure}

While the setup at ANU is capable of delivering a higher rate of numbers generated per second compared to currently accessible universal quantum computing architectures, it may still be preferable and sufficient to access trapped ion architectures, such as those provided by IonQ, or superconducting qubit systems, such as those developed by IBM\footnote{more information at ionq.com and quantum.ibm.com}, for generation through superposition of quantum states prepared using Hadamard gates.  The flow chart for the RNG using quantum hardware is shown in Figure \ref{mc_4}.

\begin{figure}[!ht]
    \centering
    \includegraphics[width=\linewidth]{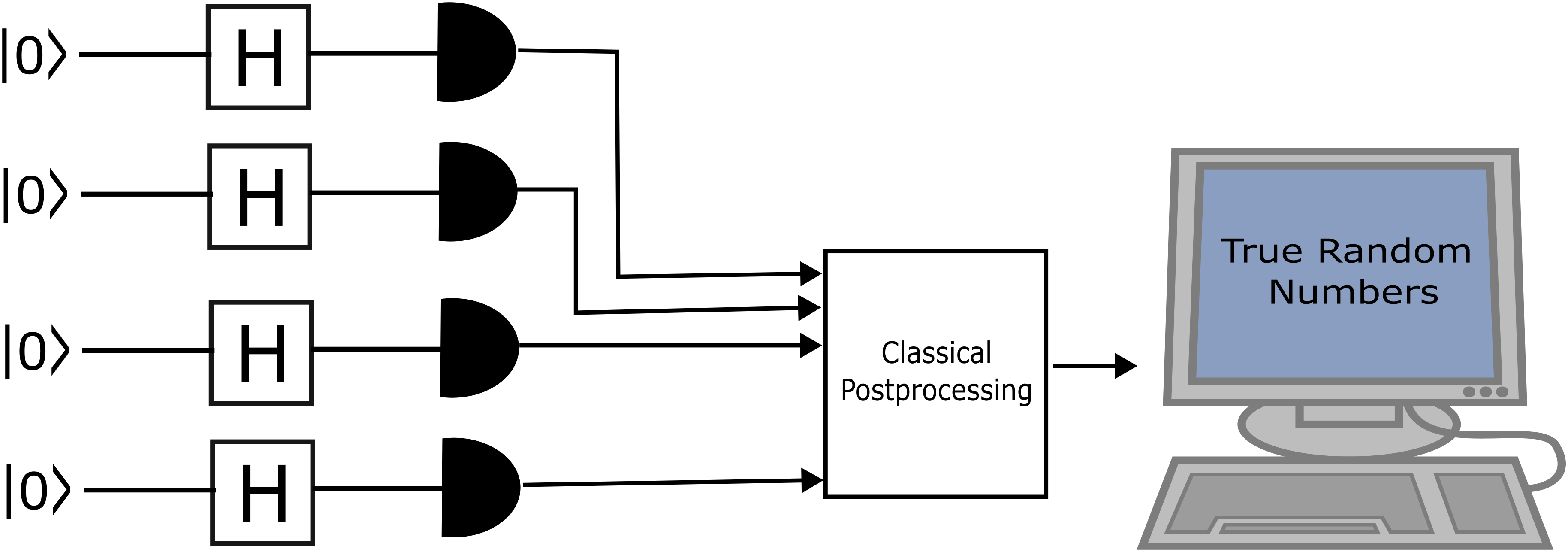}
    \caption{The architecture for quantum random number generation with the quantum circuit.}
    \label{mc_5}
\end{figure}

In these cases, the number of random samples generated would directly depend on the number of shots made by the machine.
\subsection{Benchmarking protocol for validation}
Benchmarking random number generators (RNGs) involves evaluating their performance and quality. This can be done by examining their statistical properties, speed, and suitability for various applications.  There are several tests for checking randomness for RNG, some of the statistical tests for randomness are

\begin{itemize}
\item Uniformity: The numbers should be evenly distributed across the expected range.  This can be quantified by calculating the chi-square distribution of the generated random numbers.  

\item Independence: The numbers should not be correlated. This can be quantified by calculating the autocorrelation of the generated random numbers.

\item Randomness: The entropy of the distribution can be calculated to find the randomness of the distribution.  This can be quantified by calculating the Von-Neumann entropy of the generated numbers.

\end{itemize}

\begin{table}[h]
  \centering
  \begin{tabular}{|c|c|c|}
    \hline
    Method & Von-Neumann Entropy  &Kolmogorov–Smirnov test\\ \hline
    IBM QRNG& \textbf{21.6585} &0.01430\\ \hline
    Pseudo RNG& 21.6527 &\textbf{0.00030}\\ \hline
    ANU QRNG& 21.6530 &0.00040\\ \hline
  \end{tabular}
  \caption{Validation results for the three methods of Random Number Generation}
  \label{tab:entropy}
\end{table}

In our case, we tested the RNGs obtained from ANU and random numbers generated by different quantum computers, such as IBM Brisbane. In addition, the results from the IonQ quantum simulator also show interesting metrics for the generated random numbers.  
Our analysis with the QRNGs shows the generated random numbers (except IonQ hardware) have better Von-Neumann entropy (see Table \ref{tab:entropy}) compared to RNGs generated by classical computer (we used the \textit{numpy} library for pseudo random generation) and no autocorrelation between the numbers.  
For numbers generated using IonQ hardware, the generated numbers are automatically ordered by the post-processing steps, hence proper tests cannot be conducted on IonQ hardware for our use-case. 

\begin{figure}[h]
    \centering
    \begin{adjustbox}{center}
        \includegraphics[width=1.4\linewidth]{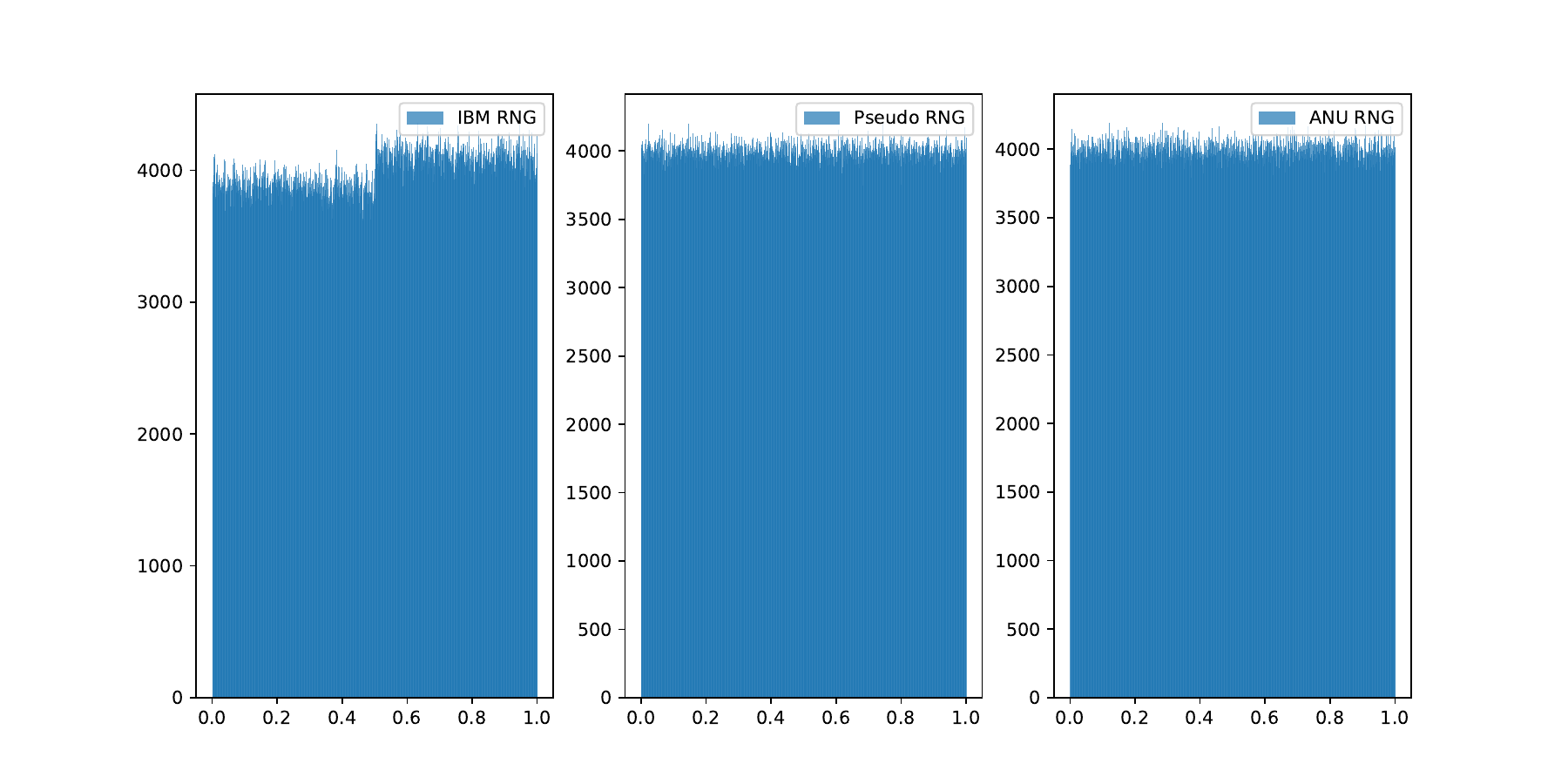}
    \end{adjustbox}
    \caption{Histograms deriving from the uniformity tests for the three methodologies implemented}
    \label{fig:uniformity_test}
\end{figure}

Moreover, we found that IBM quantum computers' qubits tend to show a slight bias towards the 0 state when measured after applying an Hadamard gate. Post-processing measures were introduced to mitigate this bias; regardless, this bias still partially affected the results of uniformity tests as can be seen in Figure \ref{fig:uniformity_test}.

\section{Results}
A series of experiments were conducted to validate and verify the proposed methods. 
As mentioned above, during the experimental phase, we compared two approaches: one that employs a dedicated setup for random number generation utilizing photonic phenomena (made accessible via ANU API protocols), and another that leverages a 127-qubit universal gate-based quantum computer for generation, IBM Brisbane \cite{ibm_brisbane}.

\begin{figure}
    \centering
    \includegraphics[angle=270, width=0.80\linewidth]{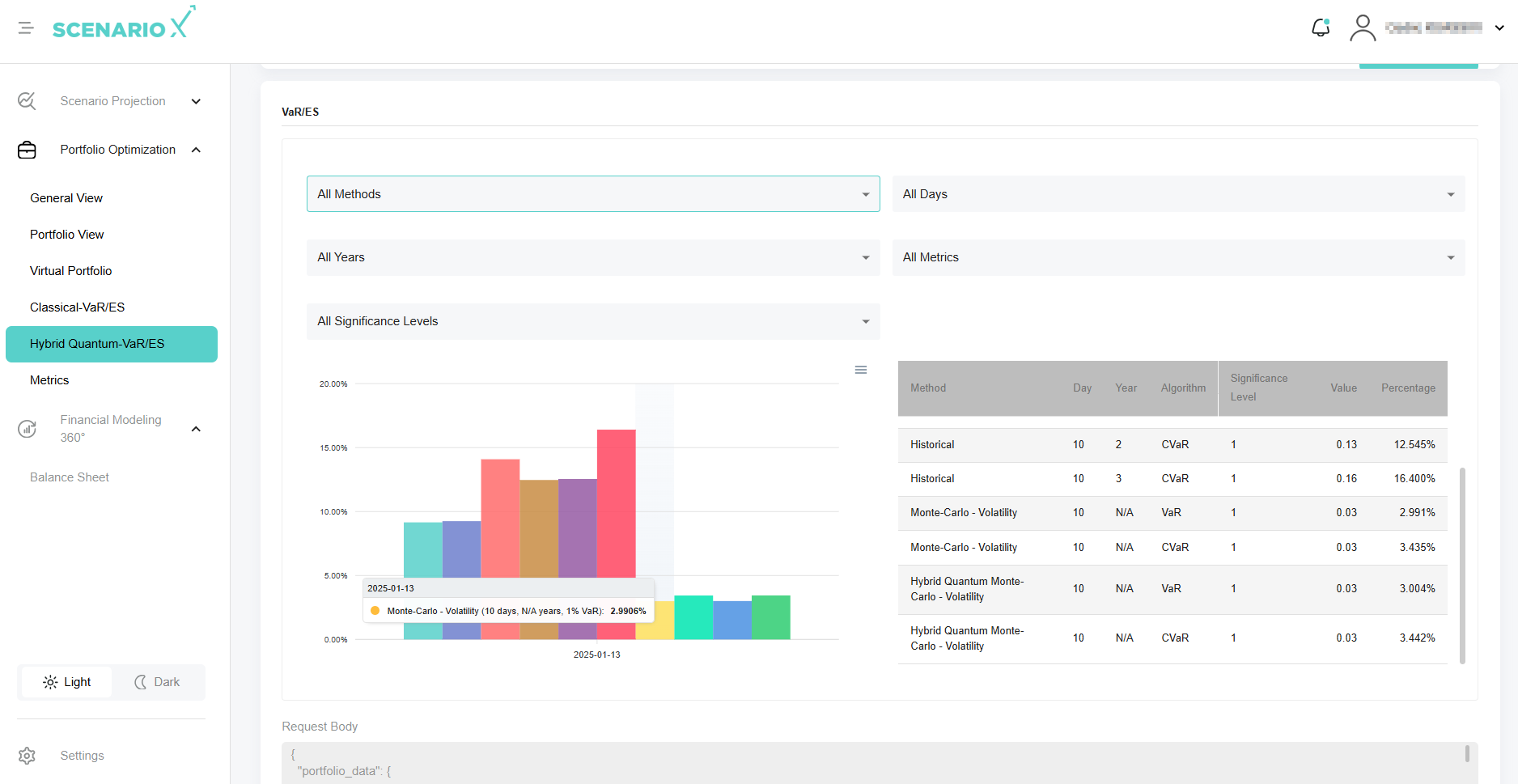}
    \caption{Scenario X SaaS Platform}
    \label{fig:platform}
\end{figure}

The results presented here are derived from a portfolio comprising $40$ different equities\footnote{The following ticker symbols are referenced as part of the analysis: AAPL, AFL, AMP, AMZN, BAC, BK, BLK, BR, CARR, CCL, CPB, CPT, CTAS, DD, EOG, FDS, FIS, GOOGL, HCA, HIG, HOLX, HON, META, MMC, MSCI, MSFT, MU, NCLH, NDAQ, NVDA, NWS, NXPI, O, ORCL, PLD, PRU, PTC, STX, TXN, UHS.} selected from indexes like the \textit{S\&P500} index. 

In order to perform the computations needed to assess the results, we used the platform \textbf{Scenario X}\footnote{www.scenario-x.ai}.
Scenario X is a cutting-edge SaaS platform designed to empower financial institutions with advanced tools for financial stress testing and decision support.
In particular, we exploited the robust risk metrics module included in the platform for calculating VaR and Expected Shortfall. This platform allows us to use:

\begin{itemize}
    \item \textbf{Historical methods} (1, 2, and 3-year horizons),
    \item \textbf{Classical Monte Carlo simulations}, and
    \item \textbf{Hybrid Quantum Monte Carlo}, combining traditional and quantum techniques for enhanced precision through Quantum Random Numbers Generation (QRNG).
\end{itemize}

These calculations are performed over a \textbf{tunable time horizon} (es. \textit{n-days = 10}) with the possibility of setting a specific \textbf{ significance level}.
Moreover, users can construct custom portfolios by selecting equities and assigning specific weights, enabling tailored risk analysis reflective of their exposures. The module provides detailed VaR and ES outputs for all methodologies, delivering actionable insights to support strategic risk management.  
A partial view of the platform dashboard is available in Figure \ref{fig:platform}.

For our experiments, the weights are initialized randomly and the two variants of the Monte Carlo process are then employed to estimate the Value at Risk (VaR) with a 2-day horizon and $2\cdot10^6$ simulated paths. 
The significance level was set to $1\%$.

\begin{figure}
    \centering
    \includegraphics[width=1.1\linewidth]{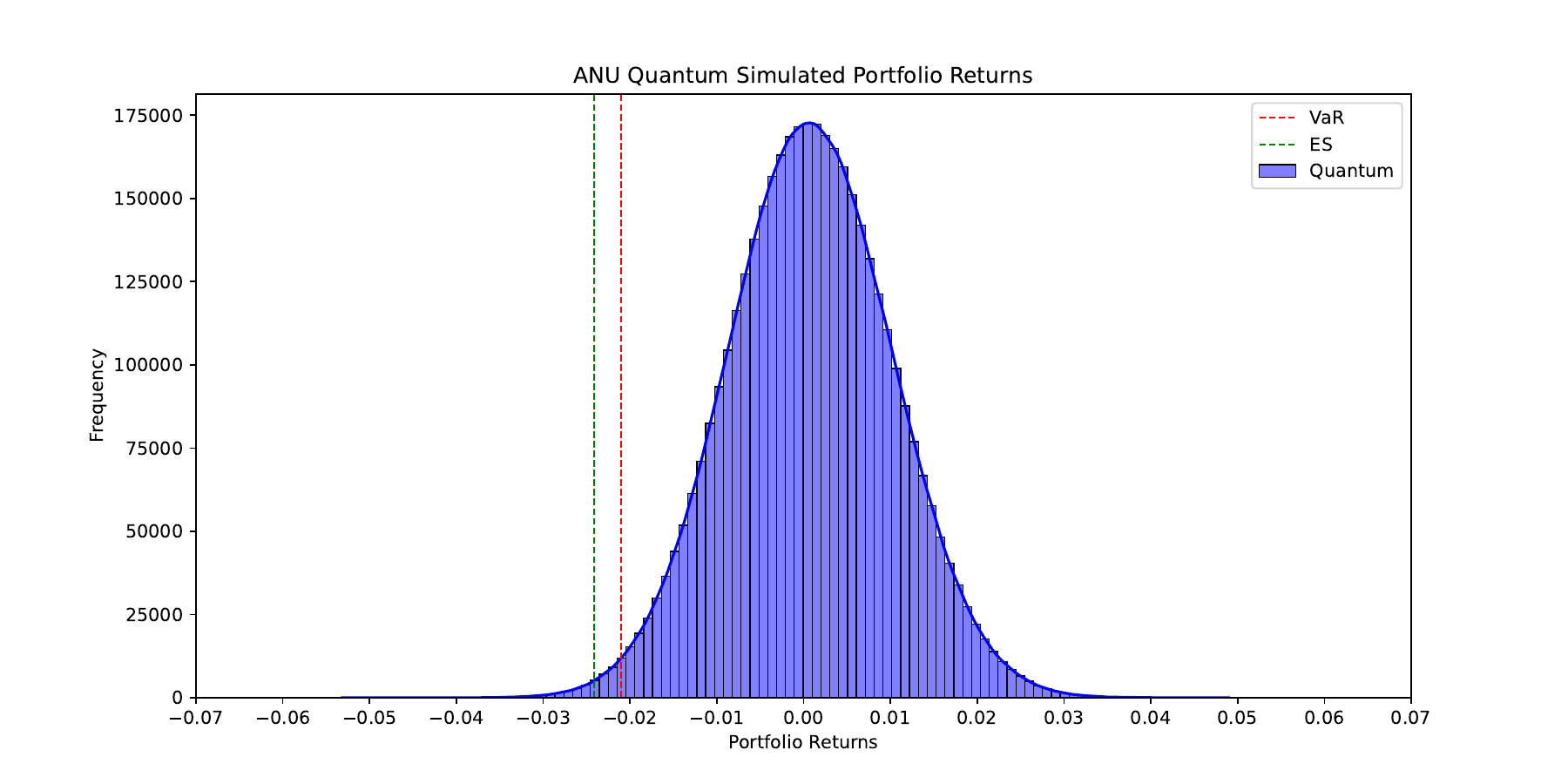}
    \caption{Returns simulated using the random numbers provided by ANU.}
    \label{fig:anu_quantum_returns}
\end{figure}

\begin{figure}
    \centering
    \includegraphics[width=1.1\linewidth]{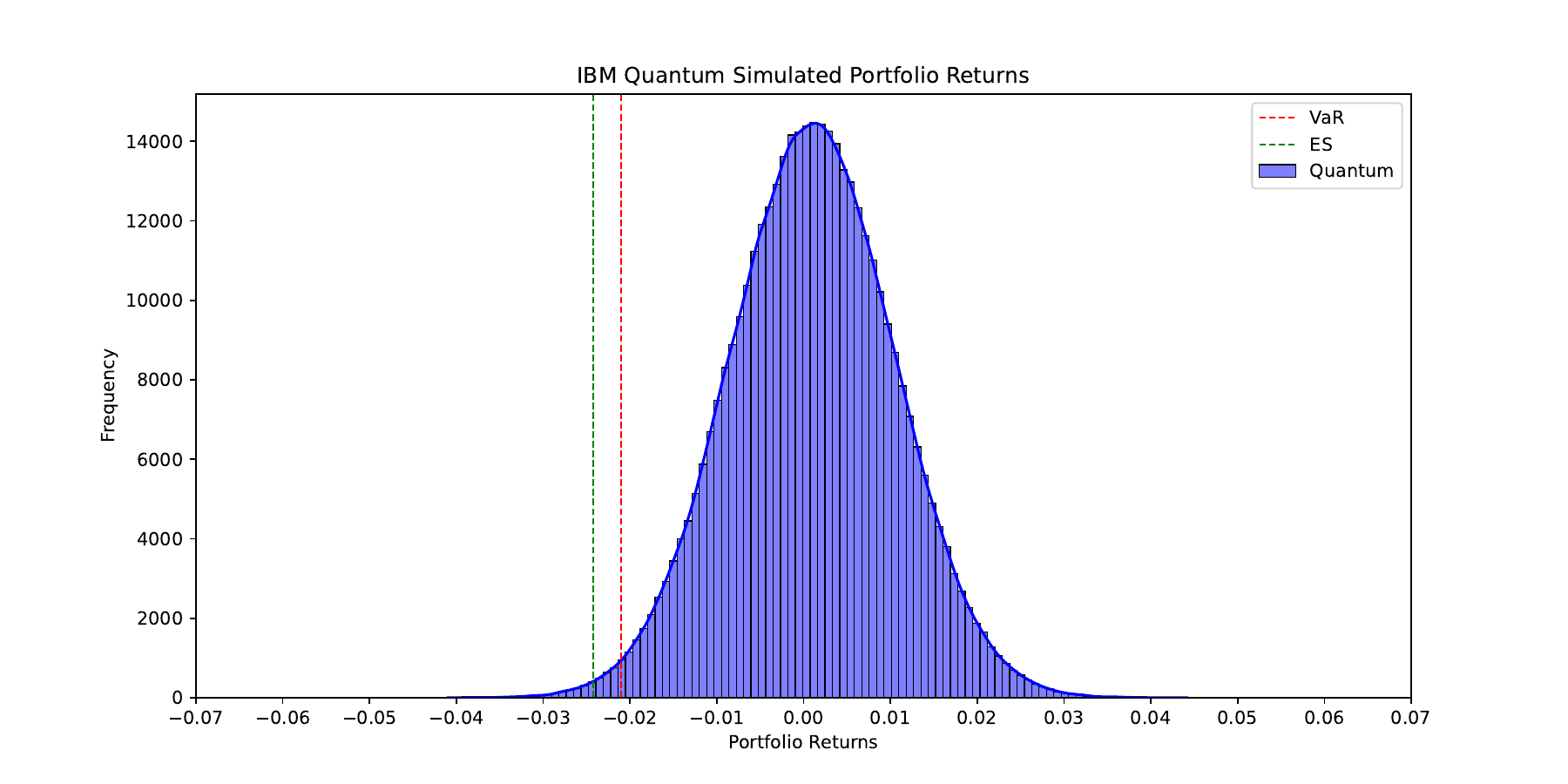}
    \caption{Returns simulated using the random numbers provided by IBM Brisbane.}
    \label{fig:ibm_quantum_returns}
\end{figure}

The results obtained using ANU-generated numbers are available in Table \ref{tab:anu_results}, while those obtained using the IBM QPU can be seen in Table \ref{tab:ibm_results}. The reconstructed distributions for both methodologies can be observed in Figure \ref{fig:anu_quantum_returns} and \ref{fig:ibm_quantum_returns}.

\begin{table}[h]
  \centering
  \begin{tabular}{lrr}
    \toprule
     & VaR & CVaR \\ \midrule
    Classical MC & 2.1069 \% & 2.4242 \% \\ 
    Quantum-enhanced & 2.1033 \% & 2.4153 \% \\ \bottomrule
  \end{tabular}
  \caption{Results using the ANU setup.}
  \label{tab:anu_results}
\end{table}

\begin{table}[h]
  \centering
  \begin{tabular}{lrr}
    \toprule
     & VaR & CVaR \\ \midrule
    Classical MC & 2.1066 \% & 2.4217 \% \\ 
    Quantum-enhanced & 2.0881 \% & 2.4087 \% \\ \bottomrule
  \end{tabular}
  \caption{Results using IBM Brisbane quantum computer.}
  \label{tab:ibm_results}
\end{table}

\subsection{Assessing Precision Improvements}
In the quest to determine whether true random numbers present a precision advantage over their pseudo random counterparts, we leveraged again the experimental setup at ANU. 

Our approach involved executing the experiments which had been previously outlined, multiple times. 
This repetitive execution was meticulously planned to provide a comprehensive analysis of the standard deviation observed in the resulting estimate. 
By doing so, we aimed to ascertain whether the natural randomness of true random numbers would lead to a narrower standard deviation, thereby indicating higher precision and reliability in experimental outcomes.

The standard deviation is a critical metric in this analysis, serving as a measure of the variability or dispersion of a set of numerical results. 
A smaller standard deviation suggests that the data points tend to be closer to the mean, implying more consistent and reliable outcomes. Conversely, a larger standard deviation indicates greater variability, which could undermine the reliability of the experimental results.

For the method employing the random numbers supplied by the quantum setup at the Australian National University (ANU), we were able to conduct five independent experiments. Each experiment involved simulating $2 \cdot 10^6$ paths. The expected value for VaR and CVaR were calculated for each experiment, the same was done also for the standard deviation among these estimates. Upon comparing the standard deviation of these quantum randomized experiments to those using pseudo random numbers, it became evident that utilizing the quantum setup resulted in a smaller standard deviation. This observation suggests an enhancement in precision and reliability when true random numbers are utilized over pseudo random numbers in this context. 
The results are reported in Table \ref{tab:precision_comparison_VaR} and \ref{tab:precision_comparison_CVaR}.
\begin{table}[h]
  \centering
  \begin{tabular}{|l|c|c|}
    \hline
    Method & Average Expected Value (VaR) & Standard Deviation \\ \hline
    Classical MC & 2.10513\% & 0.0020 \\ \hline
    Quantum - enhanced & 2.10353\% & 0.0013 \\ \hline
  \end{tabular}
  \caption{VaR - Comparison of Standard Deviation in Experimental Results Using Quantum Random Numbers vs. pseudo random Numbers.}
  \label{tab:precision_comparison_VaR}
\end{table}

\begin{table}[h]
  \centering
  \begin{tabular}{|l|c|c|}
    \hline
    Method & Average Expected Value (CVaR) & Standard Deviation  \\ \hline
    Classical MC & 2.41838\% & 0.0029 \\ \hline
    Quantum - enhanced & 2.41633\% & 0.0012 \\ \hline
  \end{tabular}
  \caption{CVaR - Comparison of Standard Deviation in Experimental Results Using Quantum Random Numbers vs. pseudo random Numbers.}
  \label{tab:precision_comparison_CVaR}
\end{table}

\section{Conclusion}
In this paper, we have presented an approach for estimating significant financial metrics within risk management by utilizing quantum phenomena for random number generation. 

The results are promising, hinting at improved accuracy with the proposed methods and slightly lower estimates (both for VaR and CVaR) using the quantum-enhanced methodology.

Furthermore, we successfully employed the Scenario X SaaS platform, demonstrating its capability and usefulness in a real-world use case at meaningful scales. This highlights how quantum technology can already benefit financial applications.


\bibliography{bibliography.bib}
\bibliographystyle{IEEEtran}

\end{document}